\documentclass[letter]{aa}  

\usepackage{graphicx}
\usepackage{txfonts}
\usepackage{lipsum}
\usepackage{subcaption}         
\usepackage{lscape}             
\usepackage{placeins}            

\usepackage{natbib}
\bibpunct{(}{)}{;}{a}{}{,} 

\usepackage[colorlinks=true, linkcolor=blue, citecolor=blue, urlcolor=blue]{hyperref}

\begin{document}

   \title{Identification of solid N$_2$O in interstellar ices using open JWST data}

   \author{V. Karteyeva
        \and R. Nakibov \and I. Petrashkevich \and M. Medvedev \and A. Vasyunin}

   \institute{Research Laboratory for Astrochemistry, Ural Federal University, Kuibysheva St. 48, Yekaterinburg 620026, Russia\\
             \email{varvara.karteeva@urfu.ru}
            }

   \date{}
 
  \abstract
   {There are only six molecules containing N-O bond that are detected in gaseous phase in interstellar medium. One of those is nitrous oxide (N$_2$O), which was searched for but not found in solid form from as early as Infrared Space Observatory (ISO) mission was launched. The observational capabilities of James Webb Space Telescope (JWST) present a possibility to identify solid interstellar N$_2$O.} 
   {We aim to identify nitrous oxide in open JWST spectra of interstellar ices towards a sample of Class 0, 0/I and flat protostars using the relevant laboratory mixtures of N$_2$O-bearing interstellar ice analogues.}
   {A set of laboratory infrared transmission spectra was obtained for the following mixtures: N$_2$O:CO$_2$=1:20, N$_2$O:CO=1:20, N$_2$O:N$_2$=1:20, N$_2$O:CO$_2$:CO=1:15:5, N$_2$O:CO$_2$:N$_2$=1:15:13 at 10--23 K. A search for N$_2$O in JWST NIRSpec spectra towards 50 protostars was performed by fitting the 4.44--4.47~$\mu$m (2250--2235 cm$^{-1}$) NN-stretch absorption band with new laboratory mixtures of N$_2$O-bearing ices.}
   {We claim the first secure identification of N$_2$O in 16 protostars. The fitting results show that N$_2$O is formed predominantly within the apolar layer of the ice mantles, rich in CO, CO$_2$ and N$_2$. The abundance of solid N$_2$O is estimated as 0.2--2.1\% relative to solid CO. We present band strengths for N$_2$O in the mixtures corresponding to the apolar layer. Also, an identification of the C-N stretch band at 4.42 $\mu$m (2260 cm$^{-1}$) is reported, which we tentatively assign to HNCO, the simplest C-N bond carrier.}
   {}


   \maketitle \nolinenumbers

\section{Introduction}
 
Nitrous oxide (N$_2$O) is one of the six N--O bond molecules detected in gaseous phase \citep{2022ApJS..259...30M}, important for the evolution of molecular complexity in space. While it was detected in the gaseous phase in various astronomical environments: Sgr B2 interstellar cloud \citep{ziurys1994detection,halfen2001evaluating}, protostellar binary IRAS 16293-2422 \citep{2018A&A...619A..28L}, G+0.693 molecular cloud \citep{rivilla2020prebiotic} and the atmosphere of Mars \citep{villanueva2013sensitive}, its solid-state counterpart --- a potential tracer of nitrogen chemistry in icy grain mantles, remains elusive. The detection of solid N$_2$O may serve as a tracer in the formation of more complex nitrogen-bearing molecules \citep{2018MNRAS.475.1819F}, act as a proxy for the presence and chemical activity of infrared-inactive molecules (N$_2$ and O$_2$) in ices \citep[e.g.][]{elsila19972140,jamieson2005investigating,pereira2018radiolysis}. Nitrous oxide could provide a direct abiotic route to species of prebiotic relevance \citep{jamieson2005investigating}, and itself was recently proposed as a potential biosignature \citep{Schwieterman_2022}.

There is a decades-long search for solid N$_2$O in interstellar ices. Initial investigations with the Infrared Space Observatory (ISO) yielded negative results \citep{1997Icar..130....1E}, highlighting the challenge of its identification. Laboratory studies, however, showed that N$_2$O is a likely product of UV-irradiated apolar ices \citep[][]{elsila19972140, moore2003infrared}. More recent experimental works also suggested that N$_2$O could be a component of cold interstellar ices and could form efficiently through ion bombardment \citep[e.g.][]{jamieson2005investigating, 2012A&A...543A.155S,pereira2018radiolysis} of apolar N$_2$-containing ices, making it a promising candidate for future observations.

N$_2$O can also be formed after ion bombardment of nitrous oxides as shown in experiments with NO$_2$:N$_2$O$_4$ mixture at 16 K and 60 K \citep{2019MNRAS.483..381F}. Finally, modeling predicts N$_2$O to be the most abundant among nitrous oxides (N$_x$O$_y$) formed under the ion bombardment of N$_2$-containing ices \citep{queiroz2025characterization}.

The unprecedented quality of James Webb Space Telescope (JWST) data reignited the search for N$_2$O. Recent studies reported tentative detections in the 4.4--4.52 $\mu$m (2272--2212 cm$^{-1}$) region \citep{nazari2024hunting} utilizing the spectrum of pure crystalline N$_2$O at 70 K, and in the 7.7 $\mu$m band towards the Ced~110~IRS4A protostar, where a spectrum of pure irradiated N$_2$O was used \citep{rocha2025ice}. There, the first N$_2$O column density estimation is provided: 8.1\texttimes10$^{16}$ cm$^{-2}$ (1.8\% relative to solid H$_2$O). Analysis of IRAS 23385+6053 also suggested a tentative assignment of a feature in the 7.7 $\mu$m region to N$_2$O \citep{Nakibov_2025} based on underfitting in this region.

However, the usage of laboratory spectra of pure or irradiated N$_2$O ices is unreliable in context of interstellar ices complexity. N$_2$O is expected to be a trace component, likely embedded in CO/CO$_2$-rich (apolar) environments. Also, the infrared band profiles and peak positions of solid species are known to shift significantly depending on the molecular environment. Currently, the following N$_2$O infrared laboratory transmission spectra are available: pure N$_2$O in 10--70 K range \citep{hudson2017infrared,gerakines2020modified}, pure N$_2$O at 16 K \citep{2009AcSpA..72.1007F}, N$_2$O:H$_2$O=1:29 at 14 K \citep{bergantini2022infrared}, and N$_2$O:CO$_2$=1:2 at 11 K \citep{pereira2018radiolysis}. Given the availability of quality observational JWST data the lack of relevant reference data becomes a primary obstacle to a secure identification. The feature in 2250--2235 cm$^{-1}$ range in the open JWST data that we explore in this Letter is a candidate for interstellar solid N$_2$O assignment based on the NN stretching mode. This region likely contains the absorption band of the N$_2$O in the apolar environment (e.g. CO$_2$, CO, N$_2$), because laboratory mixtures of N$_2$O in the polar layer do not match this range \citep[see e.g.][]{bergantini2022infrared}. There is a number of studies that support the origin of N$_2$O in apolar N$_2$-bearing ices: CO:N$_2$, N$_2$:O$_2$:CO, N$_2$:O$_2$:CO$_2$:CO \citep{elsila19972140}, CO:N$_2$ \citep{moore2003infrared}, N$_2$:CO$_2$ \citep{jamieson2005investigating}, CO:N$_2$ \citep{2012A&A...543A.155S}, N$_2$:O$_2$ \citep{2018ApJ...864...95L}.

In this Letter, we present the first reliable identification of solid N$_2$O in interstellar ices. We achieve this by combining the unparalleled sensitivity and resolution of JWST with a dedicated laboratory study of N$_2$O in the apolar environment. Our survey of 50 protostars of variable classes and masses available in the MAST database revealed clear spectroscopic signatures of N$_2$O in 16 sources, allowing us to derive its column density and establish first observational constraints on its environment in interstellar ice.

\section{Experimental setup and methods}
The transmission laboratory spectra used for N$_2$O identification were obtained using the Ice Spectroscopy Experimental Aggregate (ISEAge), a cryogenic ultra-high vacuum setup described in detail in \cite{Ozhiganov_2024}. Briefly, ISEAge setup allows for the production and study of the interstellar ice analogues. The base pressure in the chamber prior to experiments is 2$\times$10$^{-10}$ mbar, the temperature of the Ge substrate can be set and held within 6.7--305 K range. Transmission laboratory infrared (IR) spectra are obtained with the Thermo Scientific Nicolet iS50 FTIR spectrometer operated within a 4000--630~cm$^{-1}$ (2.5--15.9~$\mu$m) range with 1~cm$^{-1}$ resolution. The ices were deposited via the `background deposition' technique \citep{2011PCCP...13.8037A, 2015MNRAS.446..439F,  Rachid_et_al._2021,Rachid_et_al._2022,2024A&A...686A.236K,2025A&A...703A..89S}. The individual deposition rates are calibrated using Stanford Research Systems RGA200 quadruple mass spectrometer (QMS) in combination with IR spectroscopy via the calibration curves, which we obtained following \cite{Slavicinska_et_al._(2023)}. 

The ices were grown on a Ge substrate cooled down to 10$\pm$0.1 K. The N$_2$O deposition rate was calibrated using the band strength of 5.891\texttimes10$^{-17}$ cm for $\nu_3$ mode at 10 K taken from \cite{hudson2017infrared}. In all the experiments pure N$_2$O was introduced into the main UHV chamber at the desired fixed deposition rate through the first leak valve. Simultaneously, second gas or pre-calibrated gas mixture was introduced into the main chamber through the second independent leak valve. The QMS signal is continuously monitored during the depositions to ensure that the deposition rate is consistent with the calibration curve values. The compositions of the deposited ice mixtures were verified by examining the features in IR spectra. Co-depositions continued for 120 minutes with N$_2$O deposition rate fixed at 2.95\texttimes10$^{12}$~cm$^{-2}$~s$^{-1}$. This resulted in a total N$_2$O column density of 2\texttimes10$^{16}$~cm$^{-2}$. Afterwards, the ice was warmed up at a rate of 0.5~K per minute. During the warm up the infrared laboratory spectra were recorded every 45 seconds (averaging of 32 scans). In this Letter we present the transmission laboratory spectra of N$_2$O absorption in the apolar environment at 10~K and 23~K for the first time: N$_2$O:CO$_2$=1:20, N$_2$O:CO=1:20, N$_2$O:N$_2$=1:20, N$_2$O:CO$_2$:CO=1:15:5, N$_2$O:CO$_2$:N$_2$=1:15:13. The N$_2$O absorption bands were isolated via baseline correction for further use.

The compounds used are as follows: N$_2$O (99.999~\%, BK-Grypp), CO (99.9999~\%, Ugra-PGS), CO$_2$ (99.9999~\%, Ugra-PGS), N$_2$ (99.9999\%, Ugra-PGS). Gaseous N$_2$O, CO, CO$_2$ and N$_2$ are introduced directly into the dosing lines from the commercially acquired gas bottles.

\section{Observations and fitting procedure}
For this Letter we focused on detecting N$_2$O by its most prominent $\nu_3$ absorption band at $\sim$4.45 $\mu$m. The 7.77 $\mu$m band ($\nu_1$) is about four times less intense and was not analyzed due to insufficient signal-to-noise ratio in this region. 
 
We surveyed the MAST\footnote{MAST database: \url{https://mast.stsci.edu/portal/Mashup/Clients/Mast/Portal.html}} database for protostars observed with JWST Near-Infrared Spectrometer (NIRSpec) and compiled an initial list of 50 sources (MAST DOI: \href{https://doi.org/10.17909/tg2a-kv41}{10.17909/tg2a-kv41}). NIRspec uses G395M/G395H modes with the spectral resolution of $\sim$1000 and $\sim$2700, respectively, and covers 2.87--5.27 $\mu$m range. The science data with the level 3 pipeline calibration was used for this study \citep[calibration pipeline described by][]{Greenfield2016,Bushouse2024,Gelder2024}. We manually examined each source and searched for the apertures with detectable N$_2$O features based on our laboratory reference spectra. The search was challenged by the presence of gaseous CO emission lines that overlapped with the region of interest and, therefore, sources were classified into three groups:
\begin{enumerate}
    \item[+] Sources in which we were able to identify an aperture with N$_2$O features that had none or weak overlap with CO emission,
    \item[+] Sources in which we were able to identify an aperture with N$_2$O features that had overlap with CO emission, but the emission could be masked without causing severe distortions to the spectrum,
    \item[-] Sources in which we were not able to identify an aperture with N$_2$O features due to strong CO emission and/or low signal-to-noise ratio.
\end{enumerate}
The search yielded 16 sources (MAST DOI: \href{https://doi.org/10.17909/sg05-m334}{10.17909/sg05-m334}) in which we selected the apertures with minimal distortions in the N$_2$O region aiming for a the secure detection claim. In four objects the N$_2$O feature was recovered from gaseous CO lines. For these sources we provide an additional aperture with a clear N$_2$O feature. The intensity maps and chosen apertures are presented in Appendix \ref{appA}. The aperture parameters are presented in Table~\ref{tab:table_sources} in Appendix~\ref{appC}.

Then, the laboratory spectra were prepared for the fitting procedure by subtracting the continuum and converting them to optical depth. If required, we masked the gaseous CO emission lines. Details are presented in Appendix \ref{appB}. In some of the laboratory spectra the N$_2$O feature overlaps with a feature in 2280--2250 cm$^{-1}$ (4.39--4.44 $\mu$m) region, commonly associated with CN-stretch band, e.g. isocyanic acid (HNCO), acetonitrile (CH$_3$CN).  The CN-stretch region also contained a gaseous emission doublet \citep[H$_2$,][]{nazari2024hunting}, spanning in 2270--2265 cm$^{-1}$ range, which hindered the identification of the exact carrier. Thus, we tentatively assigned it to HNCO as it is the simplest CN-bearing molecule in this spectral region. A Gaussian function was included in the fit to cover the 2280--2250 cm$^{-1}$ feature, which resolved the overlapped bands. We also fitted Gaussian functions to estimate column densities of CO, $^{13}$CO$_2$ and OCN$^-$. The $^{12}$CO$_2$ absorption band was not analyzed due to saturation in the observational data. We also note that the profile of intense solid CO feature at 4.67~$\mu$m towards protostars can be impacted by grain shape effects \cite[see, e.g.][and references therein]{Pontoppidan_ea03}. However, this issue is not explored in this Letter.

The fitting was performed in two stages. For the first stage, we used the laboratory spectra of N$_2$O in apolar environment: mixtures with CO, CO$_2$ and N$_2$, see Fig. \ref{fig1}. The objects in our study are protostars, which are likely to have a temperature gradient in their inner regions. However, we lack detailed constraints on the physical structure of these sources. The considered observational spectra exhibit strong solid CO absorption feature at 4.67~$\mu$m. Also, the N-bearing ices we are investigating have relatively low sublimation temperatures. Finally, laboratory experiments proved the possibility of N$_2$O formation in N$_2$-containing ices at low temperatures~\citep[e.g.,][]{2018MNRAS.475.1819F}. We therefore assume that the fitted spectral features originate from cold ices residing in cold envelopes of protostars. Consequently, we use laboratory ice spectra obtained at temperatures below the sublimation point of N$_2$ (23 K) to analyze the observations. Each ice mixture spectrum is available at 35 discrete temperature points, spaced at even intervals of $\sim$0.4 K. For each source, we performed a fit at each of these 35 temperatures T$_i$ by modeling the observed spectral feature with a linear combination of laboratory spectra of different mixtures—all taken at that same T$_i$—plus an HNCO Gaussian function, minimizing $\chi^2$:

\begin{equation}\label{chi2}
    \chi^2(T_i)=\sum^{N_{obs}}_{j=1}\frac{(\tau_{obs}^j-\tau_{lab}^j(T_i))^2}{\sigma^2_j},
\end{equation}
where $\tau_{obs}$ is the optical depth of the observational spectrum, $\tau_{lab}(T_i)$ is the optical depth of a fitted linear combination of laboratory spectra and HNCO Gaussian at 2280--2250 cm$^{-1}$. $T_i$ is the temperature of the fitted combination, $N_{obs}$ is a number of points in the fitted segment of the observational spectrum and $\sigma_j$ is the point-wise standard deviation of the observational spectrum. A sliding-window smoothed curve \citep[12--16 points, locally estimated weighted scatterplot smoothing,][]{Cleveland_1979} was subtracted from the original optical depth spectrum to estimate $\sigma_i$ from the residuals. From these 35 fits across the 10--23 K range, we then selected the overall best-fit spectrum for each source.
Following \cite{Avni}, the error of temperature estimates  were obtained from $\chi^2(T_i)$ line with $\Delta\chi^2=1$ ($\alpha=68\%$). Column density errors were estimated as the best-fit uncertainty. 
The band strengths of N$_2$O in mixtures were estimated from band area measurements based on value from \cite{hudson2017infrared} (see band strengths in Table \ref{tab:table_bs}).

In the second stage the fitted spectrum was subtracted from the observational spectrum because the Gaussian in the 2280--2250 cm$^{-1}$ region overlaps with the $^{13}$CO$_2$ feature. Then, three Gaussian functions were fitted to the residual spectrum to estimate column densities of CO, OCN$^-$ and $^{13}$CO$_2$. As in the first stage, column density errors were estimated as the best fit uncertainty.

\section{Results and Discussion}
\begin{figure}[h!]
   \centering
   \includegraphics[width=\hsize]{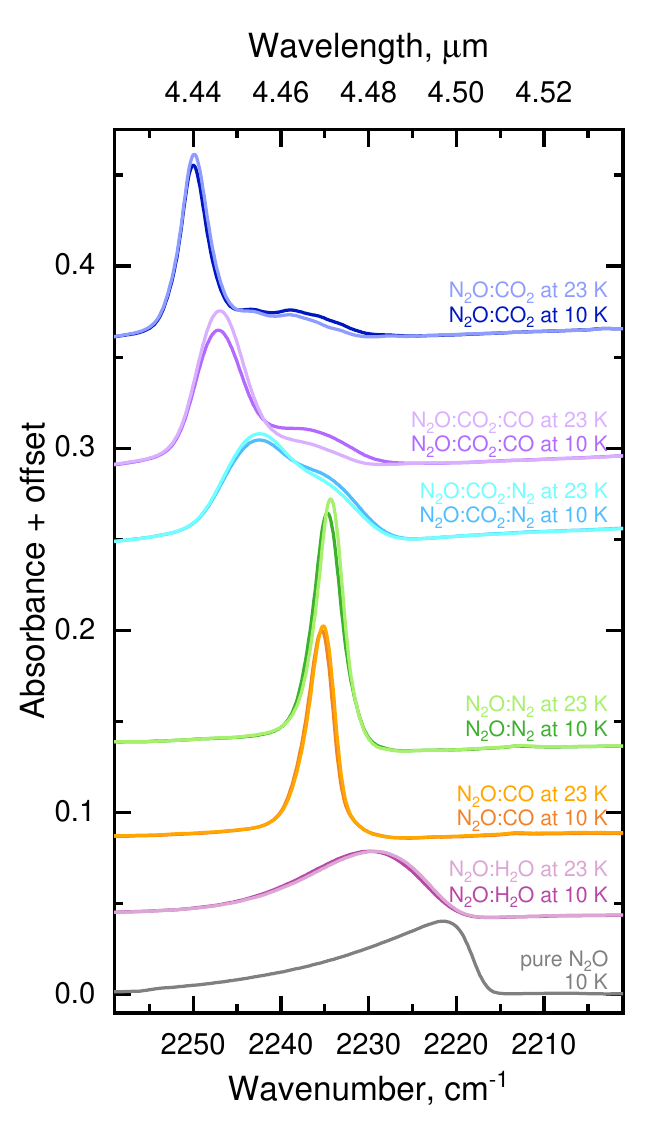}
      \caption{The $\nu_3$ mode of N$_2$O in pure N$_2$O, N$_2$O:H$_2$O=1:20 and nitrous oxide in astrochemically relevant matrices corresponding to the apolar layer of ice mantles N$_2$O:CO=1:20, N$_2$O:N$_2$=1:20, N$_2$O:CO$_2$:N$_2$=1:15:13, N$_2$O:CO$_2$:CO=1:15:5,  N$_2$O:CO$_2$=1:20. Spectra are displayed only for the lowest and the highest temperatures considered in this study.
      }
      \label{fig1}
\end{figure}

In Fig. \ref{fig1} we present the new laboratory infrared transmission spectra of N$_2$O-bearing mixtures in the apolar environment along with pure N$_2$O and N$_2$O:H$_2$O mixture at 10~K and 23~K. The spectra obtained at intermediate temperatures are not shown because the shapes of N$_2$O features vary smoothly in the 10--23~K range. The 4.45 $\mu$m ($\nu_3$) feature in apolar environment displays a significant blueshift, compared to both pure N$_2$O and N$_2$O:H$_2$O mixture. The highest blueshift of 28 cm$^{-1}$ relative to pure N$_2$O is observed for N$_2$O:CO$_2$=1:20 mixture. Ternary mixtures with CO$_2$ and CO/N$_2$ produce a broad N$_2$O feature, limited by the peak positions observed for binary mixtures with CO$_2$ and CO/N$_2$ at a ratio of 1:20. The structure of N$_2$O feature in ternary mixtures clearly indicates the presence of CO/N$_2$- and CO$_2$-associated components. Thus, the position and shape of the feature are sensitive to the ice composition. In summary, the N$_2$O absorption band in the apolar environment spans in about 2250--2235~cm$^{-1}$ range with composition-dependent line shape. 

In Fig. \ref{Fig2} the observed spectra of N$_2$O-related feature for 20 apertures in 16 protostars are shown along with best fitting results. 
As can be seen from comparison with Fig.~\ref{fig1}, the spectral feature of N$_2$O:H$_2$O mixture positioned at 2230~cm$^{-1}$ falls outside the observed features. 
All derived temperatures and column densities are listed in Table \ref{tab:table_sources}. In all observed sources, the JWST spectra were fitted with the ice mixtures corresponding to the apolar layer of the ice mantles. The derived N$_2$O column densities relative to solid CO are in the range of 0.2--2.1\%. Assuming a typical CO abundance relative to H$_2$O in a range of 10--30\%, we estimate the N$_2$O abundance with respect to H$_2$O to be in the range of 0.02--0.84\%. The 2280--2250~cm$^{-1}$ band was tentatively assigned to the C-N stretch absorption band of HNCO. It is the simplest CN-stretch mode carrier that commonly appears with N$_2$O in the irradiation experiments \citep[see e.g.][]{2018MNRAS.475.1819F}. Due to the H$_2$ emission in 2270--2265~cm$^{-1}$ range we constrained the HNCO peak position to 2260$\pm2$~cm$^{-1}$, which aligns with the 2260 cm$^{-1}$ HNCO feature reported in \cite{2018MNRAS.475.1819F}. We note that CH$_3$CN or C$_2$H$_5$CN are other possible candidates for this region \citep{nazari2024hunting}. The column densities obtained in this work for $^{13}$CO$_{2}$, OCN$^{-}$ are in agreement with values previously published in literature, see Appendix \ref{appC} for more details.

     
The detection of N$_2$O in the apolar layer is in agreement with the laboratory studies on the irradiation of N$_2$-bearing ices \citep[e.g.][]{elsila19972140,2018ApJ...864...95L}. There are both observational and chemical reasons for non-detection of N$_2$O in the $\sim$2230~cm$^{-1}$ region, that corresponds to N$_2$O embedded in H$_2$O. In observational data for this region the gaseous CO interference is stronger, which lowers the quality of the data and complicates the data selection. The chemical reason is the limited availability of N$_2$O key precursor, N$_2$, in H$_2$O-dominated ices \citep{2018ApJ...867..160H}. Interestingly, 8 of the 16 sources from the sample are the so-called HOPS sources in Orion A. Notably, this region is a subject to high UV background irradiation \citep{Peeters2024}, which further supports the irradiation-driven pathways of N$_2$O formation in interstellar ices. 

\begin{acknowledgements}
      This work is based on observations made with the NASA/ESA/CSA James Webb Space Telescope. The data were obtained from the Mikulski Archive for Space Telescopes at the Space Telescope Science Institute, which is operated by the Association of Universities for Research in Astronomy, Inc., under NASA contract NAS 5-03127 for JWST. We would like to thank Gleb Fedoseev and Vadim Krushinsky for assembling, calibrating and launching the ISEAge setup, and to Olga Russkikh for helpful discussions during manuscript revision. We thank the anonymous reviewer for their insightful comments that helped us to improve the manuscript. This research work is funded by the Russian Science Foundation via 23-12-00315 agreement.
\end{acknowledgements}

\bibliographystyle{bibtex/aa}
\bibliography{bib}

\begin{appendix}

\onecolumn
\section{Apertures}\label{appA}
\begin{figure*}[h!]
    \centering
     \resizebox{18cm}{18cm}
    {\includegraphics {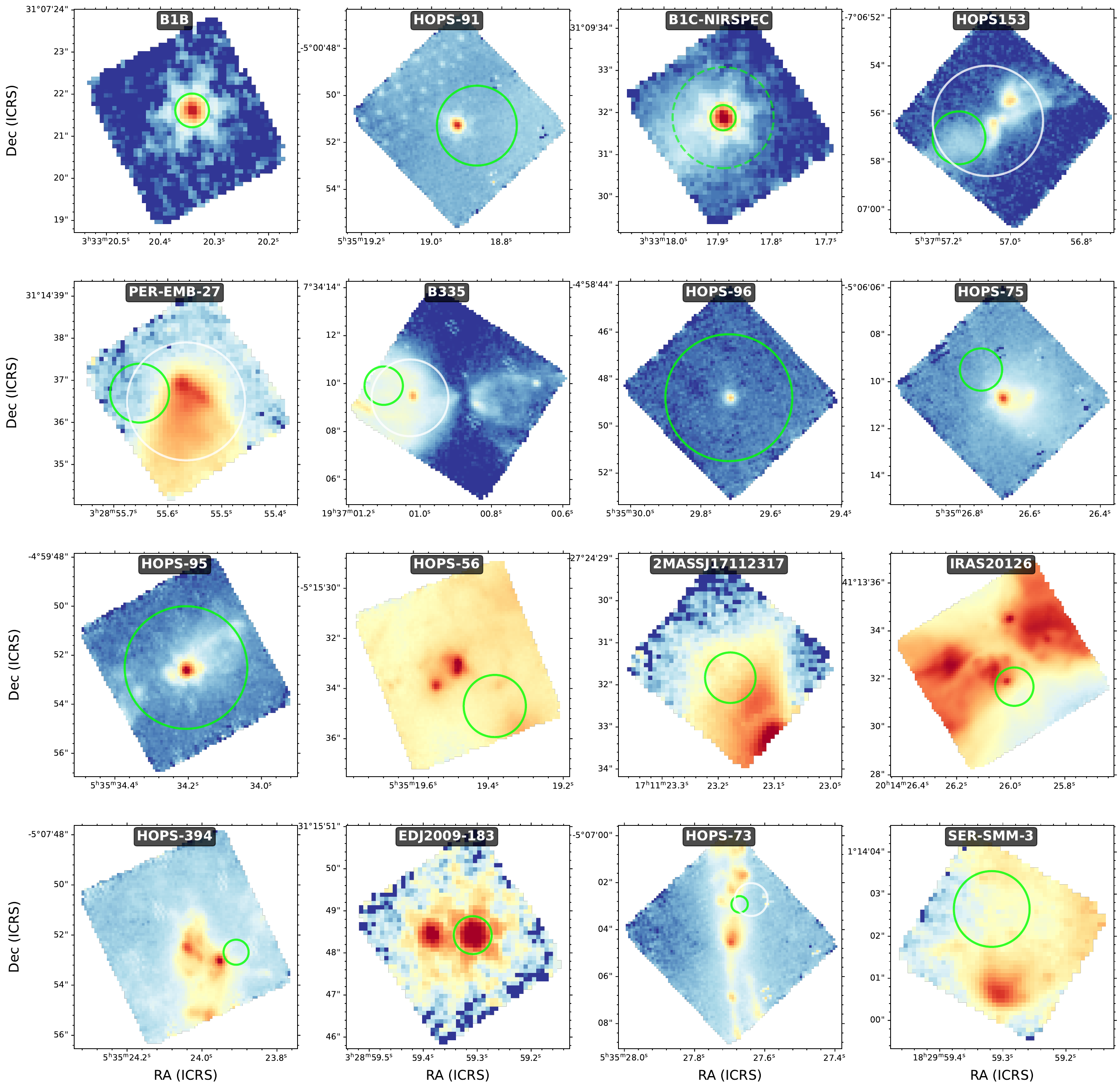}}
     
     \caption{Intensity maps at 4.4 $\mu$m with the apertures chosen for the fit. Aperture centers and diameters are listed in Table~\ref{tab:table_sources}.}
      \label{FigA}
\end{figure*}

The intensity maps at 4.4 $\mu$m and selected apertures for the sources described in this Letter are displayed in Fig. \ref{FigA}. Gaseous CO lines were masked in spectra extracted from the white apertures. Spectra extracted from the green apertures contained clean N$_2$O features as is. For B1-c a ring aperture was used to avoid gaseous lines towards the center. The data on the sources and the aperture parameters are listed in Table \ref{tab:table_sources}.

\FloatBarrier

\section{Baselines}\label{appB}
\begin{figure*}[h!]
    \centering
     \resizebox{18cm}{18cm}
    {\includegraphics {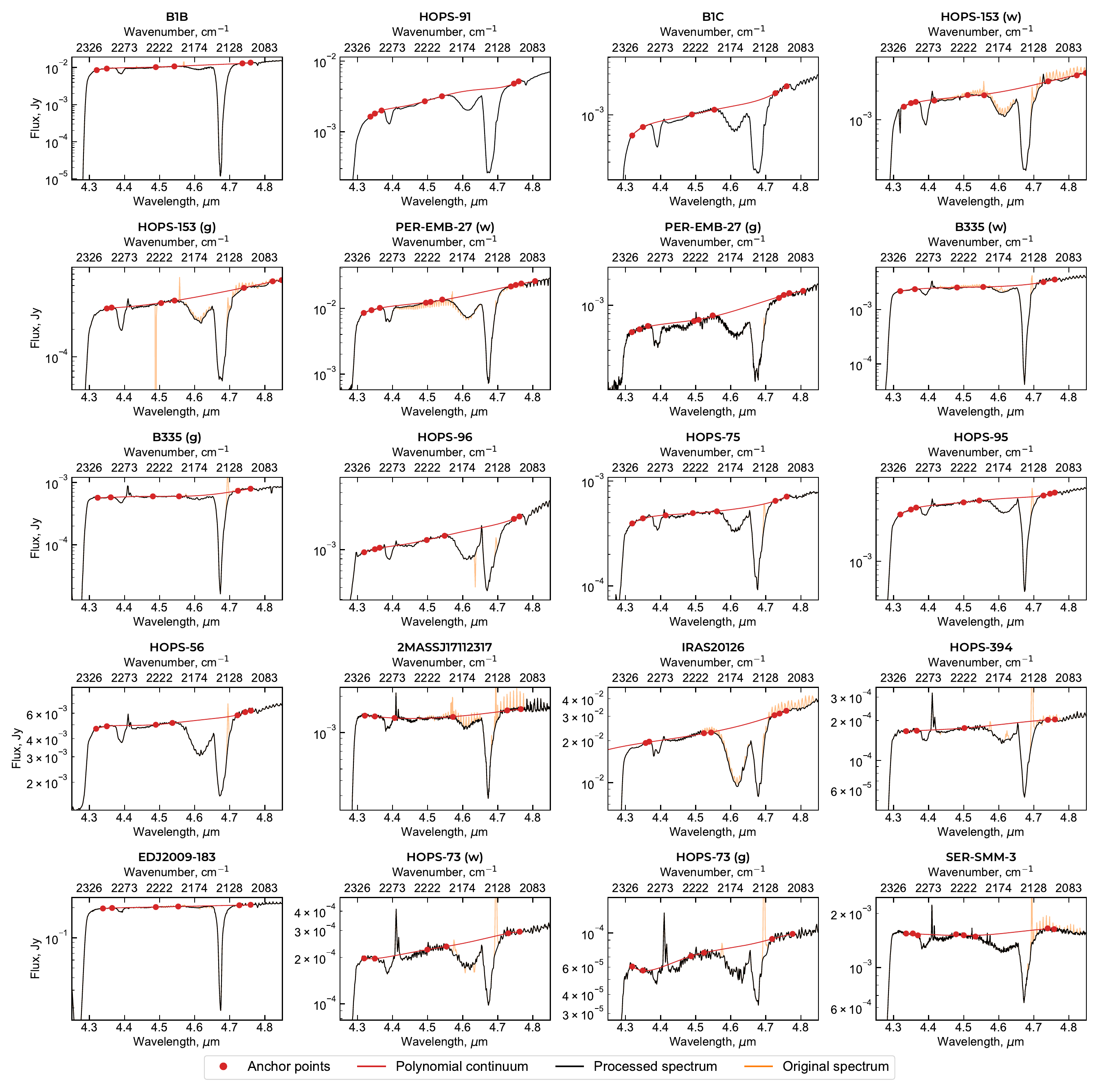}}
     
     \caption{Raw and processed observational spectra with fitted polynomial continuum. Anchor points are represented with red dots, fitted continuum with red line. Original and processed observational spectra are shown with orange and black lines, respectively.}
      \label{FigB}
\end{figure*}

This section contains original and processed spectra with masked gaseous lines. The anchor points for local continuum subtraction were selected in absorption-free regions to isolate two groups of overlapping features: $^{13}$CO$_2$, HNCO, N$_2$O and OCN$^-$, CO. Most of the points were chosen in the following ranges: 2325--2290~cm$^{-1}$, between $^{12}$CO$_2$ and $^{13}$CO$_2$; 2230--2185~cm$^{-1}$, between the proposed N$_2$O and OCN$^-$ feature; 2125--2065~cm$^{-1}$, next to the CO feature. In a few cases we included additional points outside these ranges to ensure the conservative estimation of local continuum and peak areas. A polynomial function, with a degree ranging from three to five, was selected to model the continuum. The spectra with the selected continua are shown in Fig. \ref{FigB}. 

\FloatBarrier

\section{Sources and column densities}\label{appC}

The sources and apertures' parameters along with the derived column densities are listed in Table \ref{tab:table_sources}. In this work we estimated the column density of $^{13}$CO$_{2}$  to be in the range of (1.3 -- 11)\texttimes10$^{16}$~cm$^{-2}$. This range is consistent with $^{13}$CO$_{2}$ column density range of (1.2 -- 8.1)\texttimes10$^{16}$~cm$^{-2}$ presented in \cite{2024A&A...692A.163B}, 2.3\texttimes10$^{16}$~cm$^{-2}$ value for IRAS 20126 and 2.8\texttimes10$^{16}$~cm$^{-2}$ for Per-emb 35 presented in \cite{2025ESC.....9.1992B}. \cite{2024A&A...692A.163B} also presents individual column densities for B1-b, B1-c, EDJ2009-183, SER-SMM-3, PER-EMB-27 are 2.6\texttimes10$^{16}$~cm$^{-2}$, 7.1\texttimes10$^{16}$~cm$^{-2}$, 0.58 -- 0.7\texttimes10$^{16}$~cm$^{-2}$, 2\texttimes10$^{16}$~cm$^{-2}$ and 5.9\texttimes10$^{16}$~cm$^{-2}$, respectively. These values are lower than ones derived in this work, which is explained by difference in band strengths used.

Column density values for CO obtained in this work are in (64.5 -- 464)\texttimes10$^{16}$~cm$^{-2}$ range, consistent with, for example, 6.6\texttimes10$^{17}$~cm$^{-2}$ for Ced 110 IRS4B: \citep{rocha2025ice} and (20 -- 365)\texttimes10$^{16}$~cm$^{-2}$ for Cha I \citep{2025NatAs...9..883S}. \cite{2024A&A...692A.163B} also presents individual CO column densities for B1-b, B1-c, EDJ2009-183, SER-SMM-3, PER-EMB-27: 410\texttimes10$^{16}$~cm$^{-2}$, 
500\texttimes10$^{16}$~cm$^{-2}$, 
(94 -- 100)\texttimes10$^{16}$~cm$^{-2}$, 
120\texttimes10$^{16}$~cm$^{-2}$ and 
340\texttimes10$^{16}$~cm$^{-2}$, respectively. These values are similar with the ones presented in this Letter, given the difference in chosen apertures.
Derived OCN$^{-}$ values are in range of (0.6 -- 14.6)\texttimes10$^{16}$~cm$^{-2}$, in agreement with 2.6\texttimes10$^{16}$~cm$^{-2}$ for IRAS 16253,
5.0\texttimes10$^{16}$~cm$^{-2}$ for B335, 
6.8\texttimes10$^{16}$~cm$^{-2}$ for HOPS 153, 
16\texttimes10$^{16}$~cm$^{-2}$ for HOPS 370 \citep{nazari2024hunting}, 
and 9.2\texttimes10$^{16}$~cm$^{-2}$ for Ced 110 IRS4A \citep{rocha2025ice}. To the best of our knowledge, there are currently no HNCO estimates in the literature. The band strengths for N$_2$O at 10 K in various environments were calculated based on 10 K band strength value from \cite{hudson2017infrared}. The values obtained are listed in the Table \ref{tab:table_bs}.

\begin{table*}[h!]
\caption{
Parameters of the studied objects and quantities derived from the fit.
}

\label{tab:table_sources}
\centering
\begin{tabular}{ccccccccccc}
\hline\hline 
Source & Class & Coordinates & D & T (K) &\multicolumn{5}{c}{Column densities (N), $\times10^{16}$ cm$^{-2}$} &  
R\\

                                            &
                                            &
                                            &
$^{\prime\prime}$&
N$_2$O &
N$_2$O         & 
HNCO              & 
$^{13}$CO$_{2}$ &
OCN$^{-}$         &
CO   & 
$\%$\\
\hline

B1-b$^a$           & 0  & $3^h33^m20.3415^s$ & 0.8 & $12^{+6}_{-2}$ & 1.1$\pm$0.1 & 1.3$\pm$0.1 & 5.0$\pm$1.0 & 4.7$\pm$1.0 & 464$\pm$7 & 0.2 \\
&& $31^\circ7^\prime21.6139^{\prime\prime}$  &&&&&&&\\
HOPS-91$^b$        & 0  & $5^h35^m18.8710^s$ & 3.4 & $23^{+0}_{-1}$ & 0.8$\pm$0.2 & 1.2$\pm$0.2 & 7.0$\pm$1.0 & 10.0$\pm$0.8 & 387$\pm$8 & 0.2 \\
&& $-5^\circ0^\prime51.2845^{\prime\prime}$ &&&&&&&\\
B1-c$^a$           & 0/I& $3^h33^m17.8906^s$ & 0.6 & $23^{+0}_{-4}$ & 2.2$\pm$0.2 & 2.1$\pm$0.3 & 11.0$\pm$0.9 & 13.3$\pm$0.7 & 385$\pm$7 & 0.6 \\
&&$31^\circ9^\prime31.8779^{\prime\prime}$  & 2.4$^g$ &&&&&&\\
HOPS-153$^b$ (w)$^h$   & 0  & $5^h37^m57.0635^s$ & 4.6 & $18^{+2}_{-2}$ & 1.6$\pm$0.3 & 0.4$\pm$0.2 & 8.3$\pm$0.7 & 10.1$\pm$0.6 & 261$\pm$6 & 0.6 \\
&& $-7^\circ6^\prime56.2890^{\prime\prime}$ &&&&&&&\\
HOPS-153$^b$ (g)   & 0  & $5^h37^m57.1443^s$ & 2.2 & $12^{+11}_{-2}$ & 1.0$\pm$0.2 & 0.6$\pm$0.2 & 7.9$\pm$0.7 & 11.2$\pm$0.6 & 296$\pm$6 & 0.4 \\
&& $-7^\circ6^\prime56.9967^{\prime\prime}$ &&&&&&&\\
PER-EMB-27$^a$ (w) & 0  & $3^h28^m55.5662^s$ & 2.8 & $11^{+4}_{-1}$ & 1.3$\pm$0.1 & 1.9$\pm$0.1 & 8.1$\pm$0.7 & 14.6$\pm$0.5 & 295$\pm$4 & 0.4 \\
&& $31^\circ14^\prime36.4986^{\prime\prime}$ &&&&&&&\\
PER-EMB-27$^a$ (g) & 0  & $3^h28^m55.6521^s$ & 1.4 & $11^{+2}_{-1}$ & 4.4$\pm$0.3 & 2.9$\pm$0.4 & 8.9$\pm$0.5 & 13.1$\pm$0.4 & 261$\pm$4 & 1.7 \\
&& $31^\circ14^\prime36.6958^{\prime\prime}$ &&&&&&&\\
B335$^c$ (w)       & 0  & $19^h37^m1.0285^s$ & 3.2 & $23^{+0}_{-1}$ & 1.3$\pm$0.1 & -- & 4.0$\pm$1.0 & 4.2$\pm$1.0 & 284$\pm$8 & 0.4 \\
&&$7^\circ34^\prime9.4052^{\prime\prime}$ &&&&&&&\\
B335$^c$ (g)       & 0  & $19^h37^m1.1022^s$ & 1.6 & $23^{+0}_{-1}$ & 0.7$\pm$0.1 & -- & 3.0$\pm$1.0 & 2.4$\pm$0.9 & 236$\pm$7 & 0.3 \\
&& $7^\circ34^\prime9.9127^{\prime\prime}$ &&&&&&&\\
HOPS-96$^b$        & 0  & $5^h35^m29.7202^s$ & 5.4 & $23^{+0}_{-1}$ & 1.6$\pm$0.3 & 0.3$\pm$0.1 & 4.0$\pm$1.0 & 12.6$\pm$1.0 & 212$\pm$10 & 0.8 \\
&& $-4^\circ58^\prime48.7875^{\prime\prime}$ &&&&&&&\\
HOPS-75$^b$        & 0  & $5^h35^m26.7402^s$ & 1.8 & $23^{+0}_{-3}$ & 1.2$\pm$0.5 & 0.7$\pm$0.5 & 4.2$\pm$0.6 & 7.8$\pm$0.4 & 193$\pm$4 & 0.6 \\
&& $-5^\circ6^\prime9.4794^{\prime\prime}$  &&&&&&&\\
HOPS-95$^b$        & 0  & $5^h35^m34.2059^s$ & 5   & $23^{+0}_{-3}$ & 0.7$\pm$0.2 & 0.2$\pm$0.1 & 4.0$\pm$1.0 & 6.8$\pm$0.8 & 178$\pm$6 & 0.4 \\
&&$-4^\circ59^\prime52.5012^{\prime\prime}$  &&&&&&&\\
HOPS-56$^b$        & 0  & $5^h35^m19.3829^s$ & 2.48 & $22^{+1}_{-6}$ & 1.0$\pm$0.1 & 0.6$\pm$0.1 & 4.2$\pm$0.4 & 10.4$\pm$0.4 & 162$\pm$3 & 0.6 \\
&& $-5^\circ15^\prime34.6982^{\prime\prime}$ & &&&&&&\\
2MASSJ17112317$^e$ & 0/I& $17^h11^m23.1784^s$ & 1.2 & $23^{+0}_{-1}$ & 1.2$\pm$0.1 & -- & 3.2$\pm$0.4 & 3.5$\pm$0.3 & 124$\pm$2 & 1.0 \\
&& $-27^\circ24^\prime31.8296^{\prime\prime}$ &&&&&&&\\
IRAS20126$^d$      & 0/I& $20^h14^m25.9868^s$ & 1.6 & $23^{+0}_{-10}$ & 1.0$\pm$0.2 & 0.3$\pm$0.1 & 3.7$\pm$0.4 & 17.9$\pm$0.3 & 130$\pm$2 & 0.8 \\
&& $41^\circ13^\prime31.6776^{\prime\prime}$ &&&&&&&\\
HOPS 394$^b$       & 0  & $5^h35^m23.9090^s$  & 1   & $20^{+3}_{-10}$ & 1.0$\pm$0.4 & 0.7$\pm$0.6 & 2.5$\pm$0.6 & 5.8$\pm$0.6 & 128$\pm$5 & 0.8 \\
&&$-5^\circ7^\prime52.6847^{\prime\prime}$  &&&&&&&\\
EDJ2009-183$^a$    & flat & $3^h28^m59.3083^s$ & 0.9 & $23^{+0}_{-2}$ & 0.19$\pm$0.02 & -- & 1.3$\pm$0.3 & 0.6$\pm$0.3 & 110$\pm$2 & 0.2 \\
&&$31^\circ15^\prime48.4229^{\prime\prime}$  &&&&&&&\\
HOPS-73$^b$ (w)    & 0  & $5^h35^m27.6386^s$ & 1.4 & $11^{+12}_{-1}$ & 2.0$\pm$0.5 & 0.8$\pm$0.4 & 3.7$\pm$0.6 & 7.7$\pm$0.5 & 95$\pm$4 & 2.1 \\
&&$-5^\circ7^\prime2.7197^{\prime\prime}$  &&&&&&&\\
HOPS-73$^b$ (g)    & 0  & $5^h35^m27.6716^s$ & 0.7 & $23^{+0}_{-13}$ & 2.5$\pm$0.6 & 1.2$\pm$0.8 & 2.4$\pm$0.7 & 9.6$\pm$0.8 & 83$\pm$6 & 3.0 \\
&&$-5^\circ7^\prime2.9227^{\prime\prime}$&&&&&&&\\
SER-SMM-3$^f$      & 0  & $18^h29^m59.3176^s$ & 1.8& $23^{+0}_{-1}$ & 0.5$\pm$0.4 & 1.0$\pm$0.4 & 3.0$\pm$0.3 & 7.7$\pm$0.3 & 64$\pm$2 & 0.6 \\
&&$1^\circ14^\prime2.6471^{\prime\prime}$  &&&&&&&\\
\hline
\end{tabular}

\tablefoot{
For each source listed are class, coordinates, diameter of the aperture (D), N$_2$O temperature, column density for N$_2$O, HNCO, $^{13}$CO$_{2}$, OCN$^{-}$, CO and N$_2$O abundance with respect to CO (R). Protostar classes are taken from:
\tablefoottext{a}{\cite{Tobin2016}}
\tablefoottext{b}{\cite{Furlan2016,Tobin2022}}
\tablefoottext{c}{\cite{Launhardt2013}}
\tablefoottext{d}{\cite{Chen_2016}}
\tablefoottext{e}{\cite{Riaz2009}}
\tablefoottext{f}{\cite{Ortiz-Le2018}}
\tablefoottext{g}{For B1-c the ring aperture is used, inner and outer diameters are listed.}
\tablefoottext{h}{W (white) and g (green) indicate the color of the aperture in Fig.~\ref{FigA}.}
}
\end{table*}

\begin{table*}[h!]
\caption{The list of used absorption band strength values for fit.}
\label{tab:table_bs}
\centering
\begin{tabular}{cccccccc}
\hline\hline 
Molecule & Peak Position &  Band Strength & References \\
         &      (cm$^{-1}$)          &       (cm)     &            \\
\hline
N$_2$O:CO & 2235.4 & 4.9\texttimes10$^{-17}$ & this work \\
N$_2$O:CO$_2$ & 2250 & 4.7\texttimes10$^{-17}$ & this work \\
N$_2$O:CO$_2$:CO & 2245.9& 5.2\texttimes10$^{-17}$ & this work \\
N$_2$O:CO$_2$:N$_2$ &2242.6 & 6.8\texttimes10$^{-17}$ & this work \\
N$_2$O:N$_2$ & 2234.7& 5.8\texttimes10$^{-17}$ & this work \\
HNCO & 2255 & 9.79\texttimes10$^{-17}$ & \cite{2025MNRAS.537.2918G} \\
$^{13}$CO$_2$ & 2283 & 6.8\texttimes10$^{-17}$ & \cite{Bouilloud2015} \\
OCN$^{-}$ & 2170 & 1.51\texttimes10$^{-16}$ & \cite{2025MNRAS.537.2918G}  \\
CO & 2139 & 1.12\texttimes10$^{-17}$ & \cite{Bouilloud2015} \\
\hline
\end{tabular}
\end{table*}

\newpage

\begin{figure*}[h!]

\section{Fitting results}\label{appD}
The observational spectra for the selected sample of sources, fitted N$_2$O-bearing laboratory spectra and HNCO Gaussians are shown in Fig. \ref{Fig2}.

    \centering
     \resizebox{18cm}{18cm}
    {\includegraphics {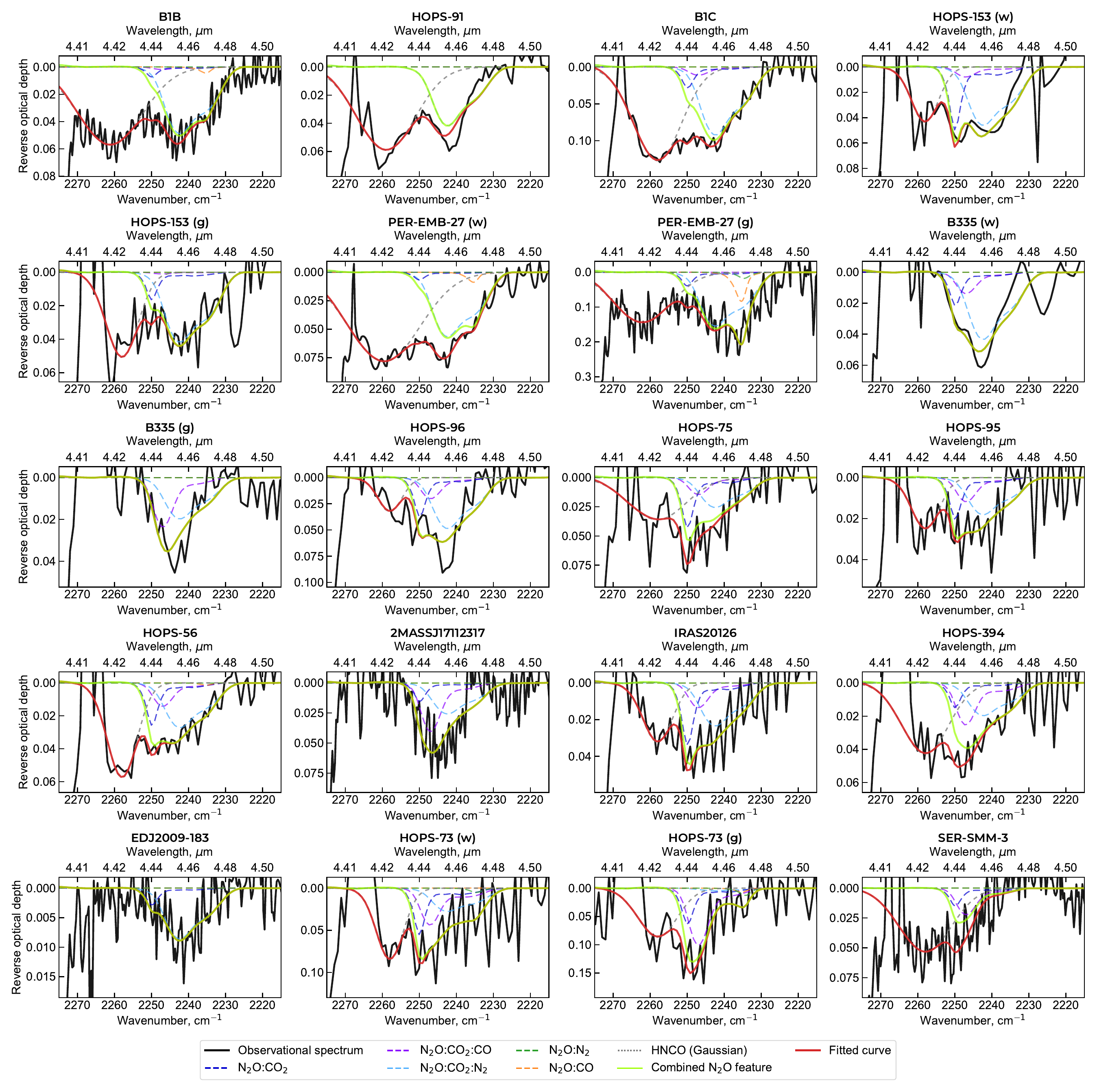}}
     
     \caption{Observational data at 4.45 $\mu$m, fitted curves and Gaussian function for HNCO. Black line represents the observational spectra, dashed --- laboratory spectra of N$_2$O in apolar environment, short dashed --- HNCO Gaussian, green --- combined N$_2$O feature, and red --- full fitted curve.}
      \label{Fig2}
\end{figure*}
\end{appendix}

\end{document}